\documentclass[nofootinbib,twocolumn,floatfix,superscriptaddress, bibliography]{revtex4-1}
\usepackage[colorlinks, linkcolor=red, anchorcolor=blue, citecolor=blue, colorlinks=true, urlcolor=blue]{hyperref}
\usepackage{amssymb}
\usepackage{amsmath}
\usepackage{amsthm}
\usepackage{mathrsfs}
\usepackage{graphicx}
\usepackage{indentfirst}
\usepackage{subfigure}
\usepackage{ulem}
\usepackage{bm}
\usepackage{xspace}
\usepackage{threeparttable}
\usepackage{gensymb}
\usepackage{multirow}
\usepackage{epstopdf}
\usepackage{breakurl}
\usepackage{lipsum}
\usepackage[figuresright]{rotating}
\usepackage{fontawesome}
\makeatother
\allowdisplaybreaks

\begin{document}

\title{Probing photon-ALP oscillations from the flat spectrum radio quasar 4C+21.35}
\author{Hai-Jun Li}
\email{lihaijun@bnu.edu.cn}
\affiliation{Center for Advanced Quantum Studies, Department of Physics, Beijing Normal University, Beijing 100875, China}

\date{\today}

\begin{abstract}

Flat spectrum radio quasar (FSRQ) is the most luminous blazar at the GeV energies.
In this paper, we probe the photon-axion-like particle (ALP) oscillation effect on the latest very-high-energy (VHE) $\gamma$-ray observations of the FSRQ 4C+21.35 (PKS~1222+216). 
The $\gamma$-ray spectra are measured by the collaborations Major Atmospheric Gamma Imaging Cherenkov Telescopes (MAGIC), Very Energetic Radiation Imaging Telescope Array System (VERITAS), and \textit{Fermi} Large Area Telescope (\textit{Fermi}-LAT), which cover two activity VHE flares of 4C+21.35 in 2010 and 2014.
We show the spectral energy distributions (SEDs) of these two phases under the null and ALP hypotheses,  and set the combined limit on the ALP parameter space.
The 95\% $\rm C.L.$ combined limit set by the FSRQ 4C+21.35 observations measured by MAGIC, VERITAS, and \textit{Fermi}-LAT in the $m_a-g_{a\gamma}$ plane is roughly at the photon-ALP coupling $g_{a\gamma} \gtrsim 8\times 10^{-12} \rm \, GeV^{-1}$ for the ALP mass $[\,2\times 10^{-10}\, {\rm eV} \lesssim m_a \lesssim 2\times 10^{-8}\, \rm eV\,]$. 
Compared with the constraint of NGC~1275 set by \textit{Fermi}-LAT, no stringent limit result is derived with the photon-ALP coupling $g_{a\gamma}$ from the FSRQ 4C+21.35, while this result could slightly broaden the ALP mass $m_a$ limit at the low-mass region.


\end{abstract}
\maketitle

\section{Introduction}

Axions are originally proposed in the Peccei-Quinn solution to solve the strong \textit{CP} problem in quantum chromodynamics (QCD) \cite{Peccei:1977ur, Peccei:1977hh, Weinberg:1977ma, Wilczek:1977pj}.
Several extensions of the Standard Model (SM) at the high energies, such as string theories \cite{Svrcek:2006yi,Arvanitaki:2009fg}, predict the axion-like particles (ALPs).
ALPs are ultralight pseudo-Nambu-Goldstone bosons (pNGBs) with the two-photon vertex $g_{a\gamma}$.
ALPs are also potential dark matter (DM) candidates \cite{Preskill:1982cy, Sikivie:2009fv, Marsh:2015xka}.
The relevant interaction of photon-ALP in the external magnetic field can be described by the effective Lagrangian \cite{Raffelt:1987im}
\begin{eqnarray}
\begin{aligned}
\mathscr{L}_{\rm ALP}&=\frac{1}{2}\partial^\mu a\partial_\mu a - \frac{1}{2}m_a^2a^2 -\frac{1}{4}g_{a\gamma}aF_{\mu\nu}\tilde{F}^{\mu\nu},
\end{aligned}
\end{eqnarray}
where $-\frac{1}{4}g_{a\gamma}aF_{\mu\nu}\tilde{F}^{\mu\nu}=g_{a\gamma}a\textbf{E}\cdot\textbf{B}$,
with the ALP field $a$, the ALP mass $m_a$, the coupling constant between the ALP and photons $g_{a\gamma}$, the (dual) electromagnetic field tensor ($\tilde{F}^{\mu\nu}$) $F_{\mu\nu}$, and the local electric and magnetic field vectors $\textbf{E}$, $\textbf{B}$.
This photon-ALP coupling in the astrophysical magnetic field environments would lead to many potentially observable photon-ALP oscillation effects \cite{DeAngelis:2007dqd, Conlon:2013txa, Giannotti:2015kwo, XENON:2020rca}.

In this work, we focus our attention on the photon-ALP oscillation effect on the very-high-energy (VHE) $\gamma$-ray photons emitted by the astrophysical sources which are far from the Milky Way, such as blazars \cite{DeAngelis:2007dqd, Hooper:2007bq}.
Blazars are radio-loud active galactic nuclei (AGNs) with the relativistic jets pointing close to our line of sight in the extragalactic VHE sky \cite{Kirk:1998kp, Urry:1995mg}.
According to the behaviors of their optical spectra, blazars can be divided into two classes, BL Lacertae (BL Lac) objects and the flat spectrum radio quasars (FSRQs) \cite{Abdo:2009iq, Padovani:2017zpf}.
The bolometric luminosity of the FSRQ is significantly greater than that of the BL Lac blazar.
In this case, we consider the photon-ALP oscillation in the blazar source region magnetic field and then further back-conversion in the Galactic magnetic field, which provides an effective way to reduce the extragalactic background light (EBL) absorption of the VHE $\gamma$-ray photon at the energy above $100\, \rm GeV$ \cite{Mirizzi:2007hr, Simet:2007sa}.
Many studies \cite{Dominguez:2011xy,DeAngelis:2011id,  Tavecchio:2012um, Horns:2012kw, Abramowski:2013oea,  Reesman:2014ova, Meyer:2014epa, Tavecchio:2014yoa,TheFermi-LAT:2016zue,  Galanti:2018myb,Galanti:2018nvl, Galanti:2018upl,  Liang:2018mqm,Zhang:2018wpc,  Libanov:2019fzq, Long:2019nrz,  Pallathadka:2020vwu, Bi:2020ths, Guo:2020kiq, Li:2020pcn, Li:2021gxs, Li:2021zms, Cheng:2020bhr, Buehler:2020qsn,Liang:2020roo, Long:2021udi} are performed to probe the effects of the photon-ALP oscillation on the VHE blazar spectra modifications or set the ALP limits in the $m_a-g_{a\gamma}$ plane.

In our previous works \cite{Guo:2020kiq, Li:2020pcn, Li:2021gxs, Li:2021zms}, we set the ALP limits from the VHE $\gamma$-ray measurements of the BL Lac blazars.
In Ref.~\cite{Guo:2020kiq}, the $\gamma$-ray observations of the BL Lac object PG 1553+113 measured by H.E.S.S. and \textit{Fermi} Large Area Telescope (\textit{Fermi}-LAT) \cite{HESS:2016btr} show the 95\% $\rm C.L.$ ALP limit at $g_{a\gamma} \gtrsim 5\times 10^{-11} \rm \, GeV^{-1}$ for the ALP mass $[\,1\times 10^{-10} \, {\rm eV} \lesssim m_a \lesssim 3\times 10^{-8}\, \rm eV\,]$.
In Ref.~\cite{Li:2020pcn}, the 4.5-year ($2008-2012$; with ten phases \cite{Bartoli:2015cvo}) BL Lac blazar Markarian 421 (Mrk~421) $\gamma$-ray observations measured by Astrophysical Radiation with Ground-based Observatory at YangBaJing (ARGO-YBJ) and \textit{Fermi}-LAT give the combined result at $g_{a\gamma} \gtrsim 2\times 10^{-11} \rm \, GeV^{-1}$ for the ALP mass $[\,5\times 10^{-10} \, {\rm eV} \lesssim m_a \lesssim 5\times 10^{-7}\, \rm eV\,]$ at 95\% $\rm C.L.$
In Refs.~\cite{Li:2021gxs, Li:2021zms}, the VHE $\gamma$-ray observations of Mrk~421 measured by Major Atmospheric Gamma Imaging Cherenkov Telescopes (MAGIC) and \textit{Fermi}-LAT in $2013-2014$ (with ten phases \cite{Acciari:2019zgl}) and 2017 (with four phases \cite{MAGIC:2021zhk}) show the almost similar combined results at 95\% $\rm C.L.$ for the ALP mass $[\,1\times 10^{-8} \, {\rm eV} \lesssim m_a \lesssim 2\times 10^{-7}\, \rm eV\,]$.

\begin{figure*}[!htbp]
\centering
  \includegraphics[width=0.75\textwidth]{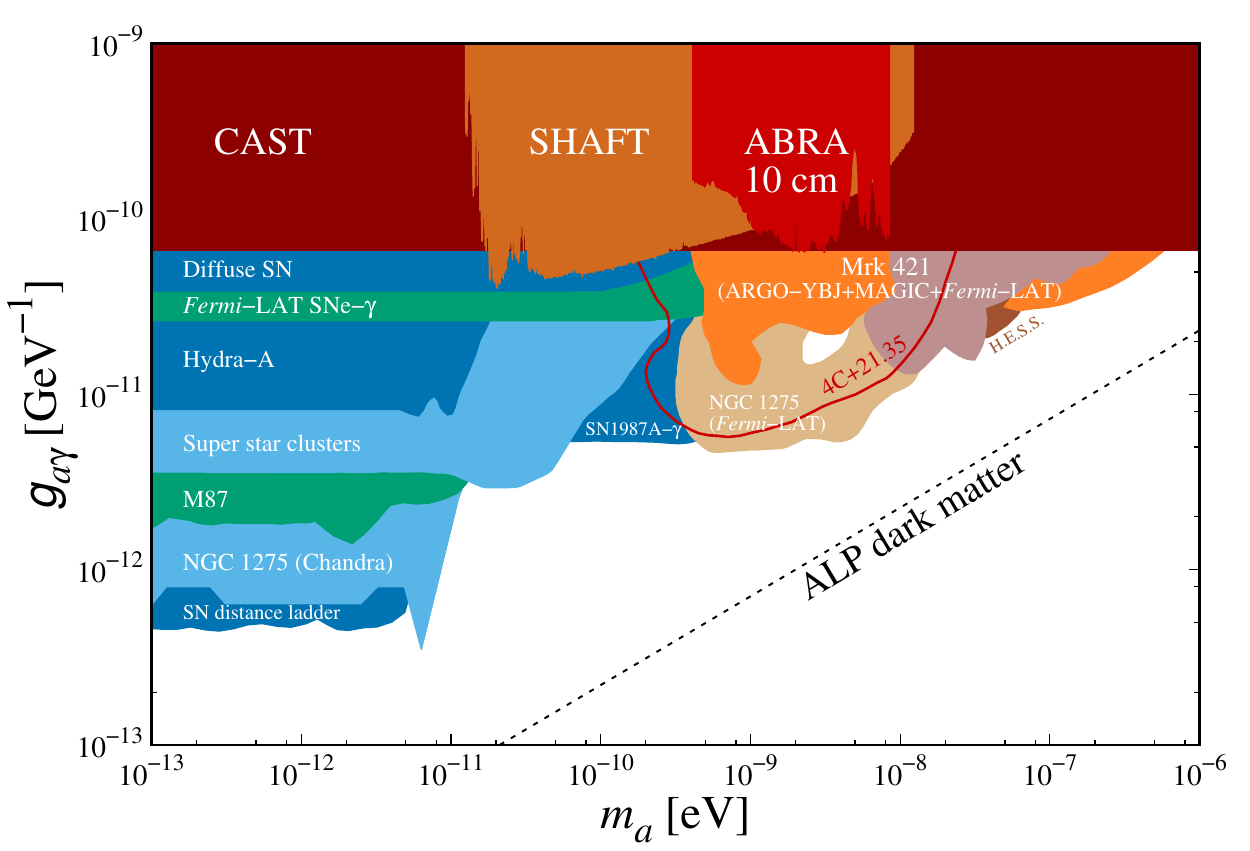}
  \caption{The 95\% $\rm C.L.$ photon-ALP limit set by the FSRQ 4C+21.35 observations measured by MAGIC, VERITAS, and \textit{Fermi}-LAT in the $m_a-g_{a\gamma}$ plane (solid red line). The other latest photon-ALP limits \cite{ciaran_o_hare_2020_3932430} at this ALP mass region are also shown. References of the data: Mrk~421 (ARGO-YBJ+MAGIC+\textit{Fermi}-LAT) \cite{Li:2020pcn, Li:2021gxs}, NGC 1275 (\textit{Fermi}-LAT) \cite{TheFermi-LAT:2016zue}, H.E.S.S. \cite{Abramowski:2013oea}, CAST \cite{Anastassopoulos:2017ftl},  SHAFT \cite{Gramolin:2020ict}, ABRA~$10\, \rm cm$ \cite{Salemi:2021gck}, Diffuse SN \cite{Calore:2021hhn}, \textit{Fermi}-LAT SNe-$\gamma$ \cite{Meyer:2020vzy}, Hydra-A \cite{Wouters:2013hua}, Super star clusters \cite{Dessert:2020lil}, M87 \cite{Marsh:2017yvc}, NGC 1275 (Chandra) \cite{Reynolds:2019uqt}, SN distance ladder \cite{Buen-Abad:2020zbd}, and SN1987A-$\gamma$ \cite{Payez:2014xsa}.
}
  \label{fig_compare_1}
\end{figure*}

We show the above ALP constraint results from the VHE $\gamma$-ray observations of the BL Lac blazars in Fig.~\ref{fig_compare_1}.
The other latest photon-ALP limits are also shown at the ALP mass $[\,1\times 10^{-13} \, {\rm eV} < m_a < 1\times 10^{-6}\, \rm eV\,]$ region.
Compared with the limit result of the NGC~1275 observation set by \textit{Fermi}-LAT \cite{TheFermi-LAT:2016zue}, we find that it would be ineffective to broaden the ALP parameter $g_{a\gamma}$ at the ALP mass $[\,1\times 10^{-10} \, {\rm eV} < m_a < 1\times 10^{-7}\, \rm eV\,]$ region with the BL Lac blazar observations, which also depends on the astrophysical magnetic field models and the uncertainties of the VHE $\gamma$-ray observations.
Additionally, only a few literatures discuss the photon-ALP oscillation effect from the FSRQ blazars \cite{Tavecchio:2012um, Mena:2013baa, Tavecchio:2014yoa,Davies:2021wqw}.
Therefore, we turn our attention to the FSRQ object and expect to derive the more stringent constraint with the photon-ALP coupling $g_{a\gamma}$ or set the ALP limit at the different ALP mass region.

Recently, the collaborations Very Energetic Radiation Imaging Telescope Array System (VERITAS) and \textit{Fermi}-LAT reported the VHE $\gamma$-ray flare observations of the FSRQ blazar 4C+21.35 in 2014 \cite{VERITAS:2021msq}.
The FSRQ 4C+21.35 (also known as PKS~1222+216), located at the redshift of $z_0 = 0.432$ \footnote{\url{http://tevcat2.uchicago.edu/sources/dDaGC0}}, was discovered at the VHE region in a flaring state by MAGIC in 2010 \cite{MAGIC:2011cjr}. 
Above two major $\gamma$-ray flares in June 2010 and February-March 2014 of 4C+21.35 are denoted as the phases Flare 2010 and Flare 2014, respectively.
In this paper, we investigate the photon-ALP oscillation effect on the VHE $\gamma$-ray observations of the FSRQ 4C+21.35 measured by MAGIC, VERITAS, and \textit{Fermi}-LAT with these two phases. 
We also set the 95\% $\rm C.L.$ combined limit on the ALP parameter ($m_a$, $g_{a\gamma}$) space.

This paper is organized as follows.
In Sec.~\ref{sec_model}, we describe the astrophysical environments model for the VHE $\gamma$-ray photons propagating from the FSRQ blazar source region to the Milky Way.
Our numerical results are presented in Sec.~\ref{sec_res}.
In Sec.~\ref{sec_con}, we comment on our results and conclude.
The statistic method is given in Appendix~\ref{app_1}.

\section{Model for the FSRQ 4C+21.35}
\label{sec_model}

In this section, we introduce the magnetic field environments model for the FSRQ 4C+21.35 with the photon-ALP beam propagating from the source region to our Earth. 
Generally, the propagation process can be divided into the blazar source region, the extragalactic space, and the Milky Way region \cite{DeAngelis:2007dqd,Hooper:2007bq}.

The main aim of this section is to derive the whole transport matrix $\mathcal{T}(s)$ of the photon-ALP beam in the above propagation process, which can be described by
\begin{eqnarray}
\mathcal{T}(s)=\mathcal{T}(s_N)_{{\rm region}-N}\times . . .\times\mathcal{T}(s_1)_{\rm region-1},
\end{eqnarray}
where $\mathcal{T}(s_i)_{{\rm region}-i}$ is the transport matrix of a single $i$-th region with the propagation distance $s_i$.
Then we can derive the final survival probability $\mathcal{P}_{\gamma\gamma}$ of the VHE $\gamma$-ray on the Earth \cite{DeAngelis:2011id,Meyer:2014epa}
\begin{eqnarray}
\mathcal{P}_{\gamma\gamma}={\rm Tr}\left(\left(\rho_{11}+\rho_{22}\right)\mathcal{T}(s)\rho(0)\mathcal{T}^\dagger(s)\right),
\label{pa}
\end{eqnarray}
with 
\begin{eqnarray}
\rho_{ii}={\rm diag}(\delta_{i1},\delta_{i2},0),
\end{eqnarray}
where $\rho(0)$ is the initial density matrix of the photon-ALP beam, and $\rho(s)=\mathcal{T}(s)\rho(0)\mathcal{T}^\dagger(s)$ is the final density matrix.

\subsection{Blazar source region}

We begin with the photon-ALP propagation process in the FSRQ blazar source region, which can also be divided into three parts, the broad line region (BLR), the blazar jet region, and the host galaxy region.
Compared with the BL Lac object, FSRQ is in a sense a more complicated version of the blazar source region.

\subsubsection{Broad line region}

For the photon-ALP beam propagates in the BLR of 4C+21.35, we consider the photon-ALP oscillation effect in the BLR magnetic field (BLRMF) and the BLR photon absorption effect due to the pair-production process \cite{Tavecchio:2012um}
\begin{eqnarray}
\gamma_{\rm VHE} + \gamma_{\rm BG} \to e^+ + e^-,
\end{eqnarray}
with the background photon $\gamma_{\rm BG}$.

The optical depth of the BLR photon absorption in this region can be described by \cite{Tavecchio:2008qv}
\begin{eqnarray}
\begin{aligned}
\tau_{\rm BLR}&=2\pi \int {\rm d}\mu \int {\rm d}\omega \int {\rm d}r(1-\mu )n(\omega,\mu,r) \\ &\times \sigma _{\gamma\gamma}(E, \omega, \mu),
\end{aligned}
\end{eqnarray}
where $\mu = \cos \theta$, $\theta$ is the scattering angle between the VHE photon $E$ and the background photon $\omega$, $r$ is the distance from the BLR center, $n(\omega,\mu,r)$ is the spectral number density of the BLR, and $\sigma_{\gamma \gamma}(E,\omega,\mu)$ is the $\gamma \gamma$ pair-production cross-section \cite{Gould:1967zzb}
\begin{eqnarray}
\begin{aligned}
\sigma&_{\gamma \gamma}(E,\omega,\mu)  = \frac{3\sigma_{\rm T}}{16} \left(1-\beta^2 \right)\\ 
&\times \left[2 \beta \left( \beta^2 -2 \right) + \left( 3 - \beta^4 \right) \, {\rm ln} \left( \frac{1+\beta}{1-\beta} \right) \right],
\end{aligned}
\end{eqnarray}
where $\sigma_{\rm T}$ is the Thompson cross-section, $\beta$ is the dimensionless parameter
\begin{eqnarray}
\beta(E,\omega,\mu) = \left[ 1 - \frac{2 \, m_e^2 \, c^4}{E \omega \left(1-\mu \right)} \right]^{1/2}. 
\end{eqnarray}
This cross-section implies that the absorption effect would become maximal for the photon energy
\begin{eqnarray}
\omega \simeq  \frac{2 \, m_e^2 \, c^4}{ E }.
\end{eqnarray}
In our analysis, the $\tau_{\rm BLR}$ distribution of the FSRQ 4C+21.35 is taken from Ref.~\cite{Tavecchio:2012um}. 

Following Ref.~\cite{Tavecchio:2012um}, we adopt the homogeneous transverse magnetic field model for the BLRMF of 4C+21.35 with the benchmark value $B^{\rm BLRMF}=0.14\, \rm G$, which is approximately consistent with the results obtained in Refs.~\cite{Kushwaha:2014jik,Kushwaha:2014qpa,Lei:2015lda}.
The radius of the 4C+21.35 BLR is adopted as $R_{\rm BLR} = 0.23\, \rm  pc$. 

\subsubsection{Blazar jet region}

Similar to the BL Lac object, we consider the photon-ALP oscillation effect in the blazar jet magnetic field (BJMF) for the FSRQ 4C+21.35 blazar jet region.
Generally, this magnetic field can be modeled as the poloidal ($B \propto r^{-2}$) and toroidal ($B \propto r^{-1}$) components \cite{Pudritz:2012xj}.
Following Refs.~\cite{Galanti:2018upl, Li:2020pcn}, we adopt the jet transverse magnetic field model $B^{\rm BJMF}(r)$ as \cite{Begelman:1984mw, Ghisellini:2009wa}
\begin{eqnarray}
B^{\rm BJMF}(r)=B_0^{\rm BJMF}\left(\frac{r}{R_{\rm BLR} }\right)^{\eta_{\rm BJMF}},
\label{bjet}
\end{eqnarray}
and the electron density distribution model $n_{\rm el}^{\rm BJMF}(r)$ as \cite{OSullivan:2009dsx}
\begin{eqnarray}
n_{\rm el}^{\rm BJMF}(r)=n_0^{\rm BJMF}\left(\frac{r}{R_{\rm BLR} }\right)^{\xi_{\rm BJMF}},
\end{eqnarray}
where $B_0^{\rm BJMF}$ and $n_0^{\rm BJMF}$ are the core magnetic field strength and the electron density at $R_{\rm BLR}$, respectively. 
Following Ref.~\cite{Tavecchio:2012um}, here we take $B_0^{\rm BJMF}=B^{\rm BLRMF}=0.14\, \rm G$, $n_0^{\rm BJMF}=10^{2} \, \rm cm^{-3}$, $\eta_{\rm BJMF}=-1$, and $\xi_{\rm BJMF}=-2$.
For the distance $r > 6.7 \rm \, kpc$, we take the magnetic field $B^{\rm BJMF}=0$.
Moreover, we introduce the photon energy transformation between the laboratory frame $E_L$ and the co-moving frame $E_j$
\begin{eqnarray}
\frac{E_L} {E_j} = \delta_{\rm D},
\end{eqnarray}
where $\delta_{\rm D}$ is the Doppler factor with the benchmark value 20 for the blazar jet region of 4C+21.35 \cite{Kushwaha:2014qpa,Lei:2015lda,Bhattacharya:2020hdg}.

\subsubsection{Host galaxy region}

FSRQ is hosted by the elliptical galaxy, the magnetic field of this region is rarely known. 
As discussed in Ref.~\cite{Tavecchio:2012um}, the photon-ALP oscillation effect in this part plays a very minor role and can be totally neglected.
Therefore, for the photon-ALP beam propagates in the host galaxy region of 4C+21.35, we do not consider the photon-ALP oscillation effect in the magnetic field.

\begin{figure*}[!htbp]
\centering
  \includegraphics[width=1\textwidth]{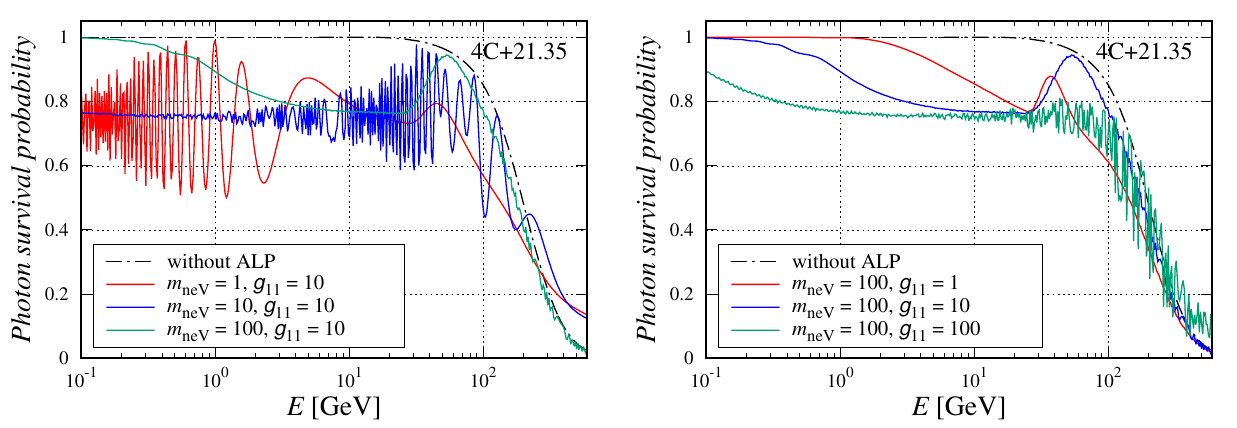}  
  \caption{The final photon survival probability distributions for the FSRQ 4C+21.35. The solid and dot-dashed lines represent the photon survival probability with/without the photon-ALP oscillation effect, respectively. Distributions for five typical ALP parameter sets (with $m_{\rm neV}$=1, 10, 100 and $g_{11}$ = 1, 10, 100) in the $m_a-g_{a\gamma}$ plane are shown. We introduce the notations $m_{\rm neV} \equiv m_a/1 \, \rm {neV} $ and $g_{11}\equiv g_{a\gamma}/10^{-11} \, \rm{GeV}^{-1}$.}
  \label{fig_pa}
\end{figure*}

Additionally, for the blazars located in the rich environment cluster, one should consider the photon-ALP oscillation effect in the turbulent inter-cluster magnetic field (ICMF) \cite{Wouters:2012qd, Meyer:2014epa}.
This magnetic field model $B^{\rm ICMF}(r)$ can be described by 
\begin{eqnarray}
B^{\rm ICMF}(r)=B_0^{\rm ICMF}\left(\frac{n_{\rm el}^{\rm ICMF}(r)}{n_0^{\rm ICMF}}\right)^{\eta_{\rm ICMF}},
\end{eqnarray}
with the electron density distribution model $n_{\rm el}^{\rm ICMF}(r)$
\begin{eqnarray}
n_{\rm el}^{\rm ICMF}(r)=n_0^{\rm ICMF}\left(1+\frac{r}{R_{\rm core}}\right)^{\xi_{\rm ICMF}},
\end{eqnarray}
where $B_0^{\rm ICMF} = \mathcal{O}(1)\, \mu \rm  G$, $n_0^{\rm ICMF}=\mathcal{O}(10^{-3})\, \rm cm^{-3}$, $R_{\rm core} =\mathcal{O}(100)\rm \, kpc $, $0.5 \lesssim \eta_{\rm ICMF} \lesssim 1$, and $\xi_{\rm ICMF}=-1$.
However, there is no definite evidence that the FSRQ 4C+21.35 considered in this work is located in the rich environment cluster. 
Thus, we do not take into account the photon-ALP oscillation effect in this ICMF.

\subsection{Extragalactic space}

For the photon-ALP beam propagates in the extragalactic space, we also consider the EBL photon absorption effect due to the pair-production process. 
The optical depth of the EBL photon absorption in this region can be described by \cite{Belikov:2010ma, Franceschini:2008tp}
\begin{eqnarray}
\begin{aligned}
\tau_{\rm EBL}&=c \int_0^{z_0} \frac{{\rm d}z}{(1+z)H(z)}\int_{E_{\rm th}}^{\infty}{\rm d}\omega\frac{{\rm d}n(z)}{{\rm d}\omega}\\
&\times\bar{\sigma}(E,\omega,z),
\end{aligned}
\end{eqnarray}
with
\begin{eqnarray}
H(z)=H_0\left[\left(1+z\right)^2\left(1+\Omega_m z\right)-z\left(2+z\right)\Omega_\Lambda\right]^{1/2},~~~
\end{eqnarray}
where $z_0$ is the redshift of the source, $E_{\rm th}$ is the threshold energy, $\bar{\sigma}(E,\omega,z)$ is the integral pair-production cross-section, ${\rm d}n(z)/{\rm d}\omega$ is the proper number density of the EBL, $H_0 \simeq 67.4\, \rm km\, s^{-1} \,Mpc^{-1}$ is the Hubble constant, $\Omega_m \simeq 0.315$ is the matter density, and $\Omega_\Lambda \simeq 0.685$ is the dark energy density \cite{ParticleDataGroup:2020ssz}.
In this work, we take the EBL model from Ref.~\cite{Franceschini:2008tp}.

Moreover, we do not consider the photon-ALP oscillation effect in the extragalactic magnetic field.
The strength of this magnetic field is very weak (current limit: $10^{-7} \, {\rm nG} \lesssim {B}_{\rm ext} \lesssim 1.7 \,  {\rm nG}$ \cite{Ade:2015cva, Pshirkov:2015tua}) and can not be explicitly measured.

\subsection{Milky Way}

For the photon-ALP beam propagates in the Milky Way, we also consider the photon-ALP oscillation effect in the Galactic magnetic field.
Following Refs.~\cite{Jansson:2012pc, Jansson:2012rt}, the Galactic magnetic field can be modeled as the disk and halo components (parallel to the Galactic plane), and the ``X-field" component (out-of-plane at the Galactic center).
One can find the latest version about this magnetic field model in Ref.~\cite{Planck:2016gdp}.

\begin{figure*}[!htbp]
\centering
  \includegraphics[width=1\textwidth]{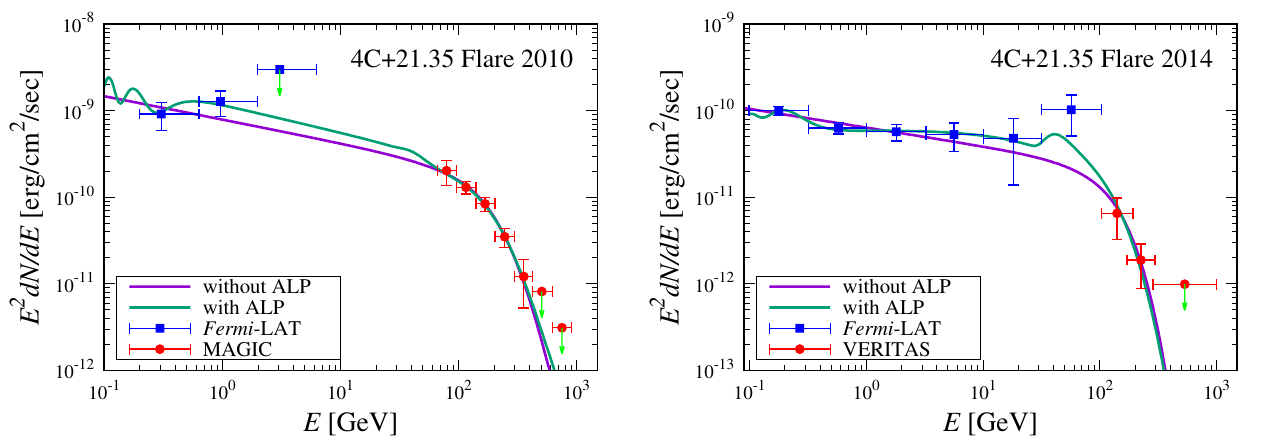}
  \caption{The best-fit $\gamma$-ray SEDs of the FSRQ 4C+21.35 under the null and ALP hypotheses with the two phases Flare 2010 (left) and Flare 2014 (right). The purple and green lines represent the spectra under the null and ALP hypotheses, respectively. The experimental data of MAGIC, VERITAS, and \textit{Fermi}-LAT are taken from Refs.~\cite{MAGIC:2011cjr, VERITAS:2021msq}. The values of the best-fit $\chi_{\rm null}^2$/$\chi_{\rm ALP}^2$ in the phases Flares 2010 and 2014 are 1.86/0.22 and 4.36/1.85, respectively.}
  \label{fig_dnde}
\end{figure*}

\section{Results}
\label{sec_res}

With the magnetic field environments model for the FSRQ 4C+21.35 considered in Sec.~\ref{sec_model}, we can derive the final survival probability of the VHE $\gamma$-ray on the Earth, which is shown in Fig.~\ref{fig_pa}.
The solid and dot-dashed lines represent the photon survival probability with/without the photon-ALP oscillation effect, respectively. 
We show the final photon survival probability distributions for five typical ALP parameter sets (with $m_{\rm neV}$=1, 10, 100 and $g_{11}$ = 1, 10, 100) in the $m_a-g_{a\gamma}$ plane.
The behavior of the photon-ALP oscillation varies dramatically with the ALP mass $m_a$ and the coupling constant $g_{a\gamma}$.

Using the final photon survival probability $P_{\gamma\gamma}$ in Eq.~(\ref{pa}), we can derive the expected VHE $\gamma$-ray spectrum
\begin{eqnarray}
\varPhi_{\rm exp} ( E ) = \mathcal{P}_{\gamma\gamma} \varPhi_{\rm int} ( E ),
\end{eqnarray}
where $\varPhi_{\rm int}(E)$ is the $\gamma$-ray intrinsic spectrum. 
Following Ref.~\cite{Acciari:2019zgl}, the VHE blazar intrinsic spectrum model can be described by a simple and smooth concave function with several free parameters.
In Ref.~\cite{Li:2021zms}, we explored the effects between the $\gamma$-ray intrinsic spectrum model selections and the ALP limits, which showed an insignificant relationship between them.
For the phase Flare 2010 of 4C+21.35, we use the power law with exponential cut-off (EPWL; three free parameters) model
\begin{eqnarray}
\frac{{\rm d}N}{{\rm d} E}= N_0\left(\frac{E}{E_0}\right)^{-\varGamma}\exp\left(-\frac{E}{E_c}\right),
\end{eqnarray}
while for the phase Flare 2014, we use the power law with super-exponential cut-off (SEPWL; four free parameters) model 
\begin{eqnarray}
\frac{{\rm d}N}{{\rm d}E} = N_0\left(\frac{E}{E_0}\right)^{-\varGamma}\exp\left(-\left(\frac{E}{E_c}\right)^d\right),
\end{eqnarray}
with the normalization constant $N_0$, the normalization energy $E_0 = 1\, \rm GeV$, the spectral index $\varGamma$, and the free parameters $E_c$ and $d$. 

Then the $\chi^2$ value can be described by
\begin{eqnarray}
\chi^2 = \sum_{i=1}^{N} \left(\frac{\varPhi_{\rm exp}(E_i) - \tilde{\phi}_i}{\delta_i}\right)^2,
\label{chi2}
\end{eqnarray}
with the spectral point number $N$, the expected $\gamma$-ray spectrum $\varPhi_{\rm exp}(E_i)$, the detected $\gamma$-ray  spectrum $\tilde{\phi}_i$, and the corresponding uncertainty of the observation $\delta_i$.
The best-fit $\gamma$-ray spectral energy distributions (SEDs) of the two phases Flares 2010 and 2014 of 4C+21.35 under the null hypothesis are shown in Fig.~\ref{fig_dnde}.

\begin{figure}[!htbp]
\centering
  \includegraphics[width=0.491\textwidth]{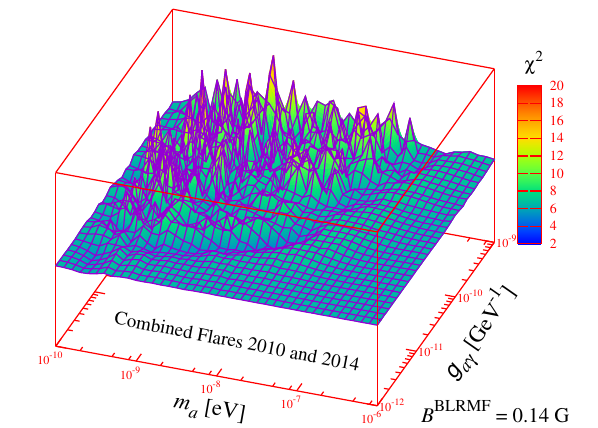}
  \includegraphics[width=0.491\textwidth]{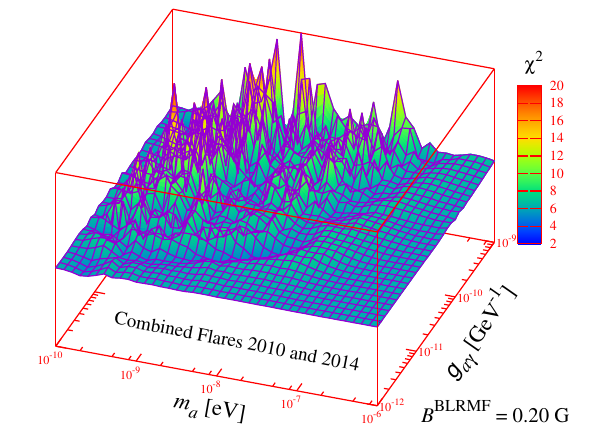}
  \caption{The $\chi_{\rm ALP}^2$ distributions in the $m_a-g_{a\gamma}$ plane for the two phases Flares 2010 and 2014 of the FSRQ 4C+21.35 combined. The top and bottom panels represent the parameter $B^{\rm BLRMF}$ $(B_0^{\rm BJMF})= 0.14 \, \rm G$ and $0.20 \, \rm G$, respectively.}
  \label{fig_3d}
\end{figure}

For each ALP parameter set in the $m_a-g_{a\gamma}$ plane, we can derive the value of the best-fit $\chi_{\rm ALP}^2$.
We also give the minimum best-fit $\gamma$-ray SEDs for the two phases of 4C+21.35 under the ALP hypothesis in Fig.~\ref{fig_dnde} for comparisons.
Following Refs.~\cite{Li:2020pcn, Li:2021gxs}, we show the result for these two phases combined together, which is essential for the multi-phase analysis.
The $\chi_{\rm ALP}^2$ distribution in the $m_a-g_{a\gamma}$ plane for the two phases Flares 2010 and 2014 of 4C+21.35 combined is shown in Fig.~\ref{fig_3d}.

Following Refs.~\cite{TheFermi-LAT:2016zue, Li:2020pcn}, in order to set the 95\% $\rm C.L.$ limit on the ALP parameter space, 400 sets of the $\gamma$-ray spectra observations produced by Gaussian samplings in the pseudo-experiments are simulated to obtain the test statistic (TS) distribution and the value of $\Delta\chi_{95\%}^2$ (see Appendix~\ref{app_1} for more details).
For the two phases Flares 2010 and 2014 of 4C+21.35 combined, we derive the value of $\Delta\chi_{95\%}^2 = 5.87$ at 95\% $\rm C.L.$ with the effective ${\rm d.o.f.}=1.92$ and the non-centrality $\lambda=0.01$ \footnote{A convenient non-central $\chi^2$ distribution calculator is available online at \url{https://keisan.casio.com/exec/system/1180573184}.}.

\begin{figure}[!htbp]
\centering
  \includegraphics[width=0.5\textwidth]{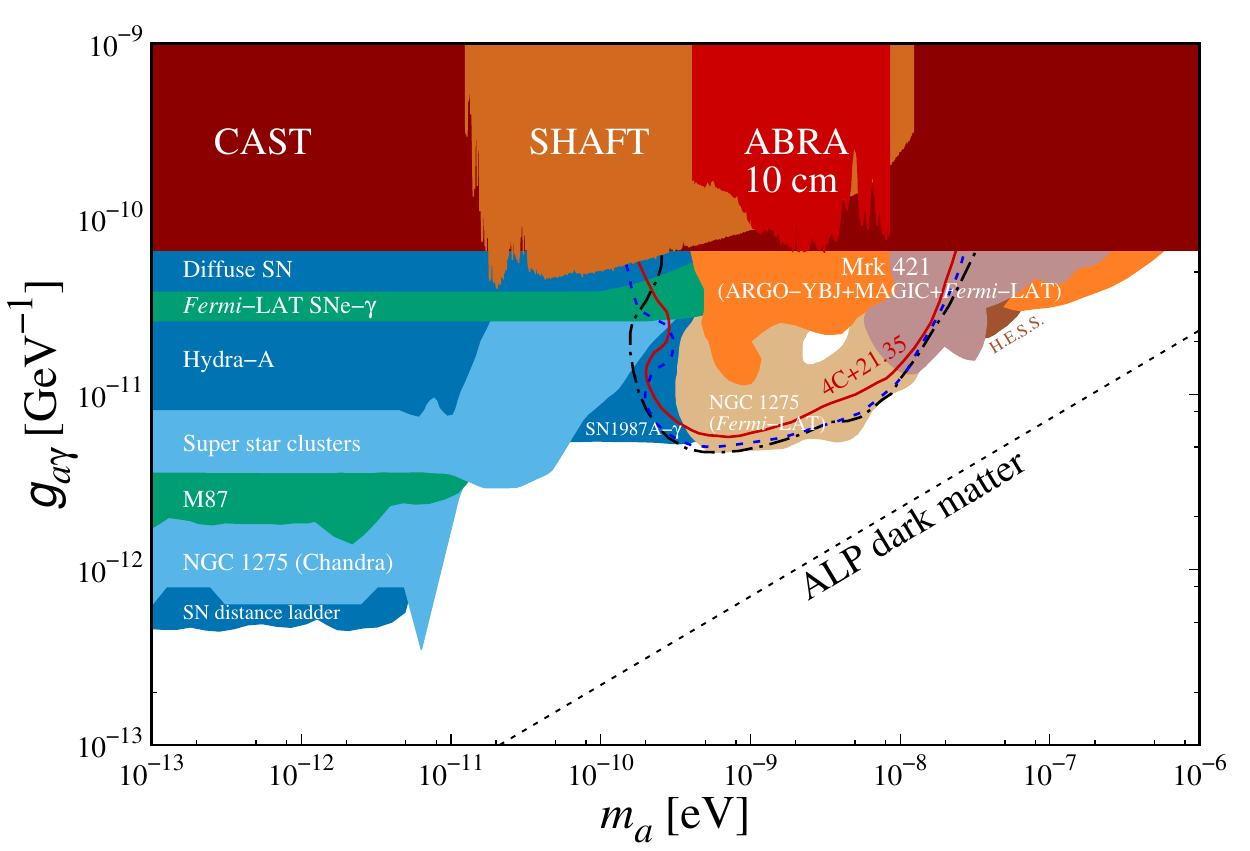}
  \caption{Same as Fig.~\ref{fig_compare_1} but for the different magnetic field environments model parameters. The dot-dashed black and dashed blue lines represent the 95\% $\rm C.L.$ combined limits set by the FSRQ 4C+21.35 observations with the parameters $B^{\rm BLRMF}$ $(B_0^{\rm BJMF})= 0.20 \, \rm G$ and $R_{\rm BLR}=0.30 \rm \, pc$, respectively. In the analysis, we just change one parameter and adopt the other parameters as the benchmark values.}
  \label{fig_compare_2}
\end{figure}

We show the 95\% $\rm C.L.$ photon-ALP combined limit set by the FSRQ 4C+21.35 observations measured by MAGIC, VERITAS, and \textit{Fermi}-LAT in Fig.~\ref{fig_compare_1} (one can also find in Fig.~\ref{fig_compare_2}) with the photon-ALP coupling at $g_{a\gamma} \gtrsim 8\times 10^{-12} \rm \, GeV^{-1}$ for the ALP mass $[\,2\times 10^{-10}\, {\rm eV} \lesssim m_a \lesssim 2\times 10^{-8}\, \rm eV\,]$. 
The other latest photon-ALP limits at the ALP mass $[\,1\times 10^{-13} \, {\rm eV} < m_a < 1\times 10^{-6}\, \rm eV\,]$ region are also shown for comparisons.
Compared with the constraint of NGC~1275 set by \textit{Fermi}-LAT \cite{TheFermi-LAT:2016zue}, we could not obtain a more stringent limit result with the photon-ALP coupling $g_{a\gamma}$ from the FSRQ 4C+21.35, while this result could slightly broaden the ALP mass $m_a$ limit at the mass $m_a \simeq 2\times 10^{-10}\, \rm eV$ region.

As discussed in Ref.~\cite{Li:2020pcn}, the magnetic field environments model parameters $B^{\rm BLRMF}$ ($B_0^{\rm BJMF}$) and $R_{\rm BLR}$ would significantly affect the final limit results and can not be completely determined in this work.
For comparison, we give the ALP limit results with the large values of these two model parameters.
In Fig.~\ref{fig_compare_2}, the 95\% $\rm C.L.$ photon-ALP combined limits set by the FSRQ 4C+21.35 observations with the parameters $B^{\rm BLRMF}$ $(B_0^{\rm BJMF})= 0.20 \, \rm G$ ($\chi_{\rm ALP}^2$ distribution is shown in Fig.~\ref{fig_3d}) and $R_{\rm BLR}=0.30 \rm \, pc$ are given, which show the slightly more stringent limits.

\section{Conclusion}
\label{sec_con}

In this paper, we have presented the photon-ALP oscillation effect on the latest VHE $\gamma$-ray observations of the FSRQ 4C+21.35 measured by VERITAS and \textit{Fermi}-LAT in 2014.
Combined with the 4C+21.35 observations measured by MAGIC and \textit{Fermi}-LAT in 2010, the 95\% $\rm C.L.$ limit on the ALP parameter space is roughly at the photon-ALP coupling $g_{a\gamma} \gtrsim 8\times 10^{-12} \rm \, GeV^{-1}$ for the ALP mass $[\,2\times 10^{-10}\, {\rm eV} \lesssim m_a \lesssim 2\times 10^{-8}\, \rm eV\,]$. 
Compared with the constraint of NGC~1275 set by \textit{Fermi}-LAT, no stringent limit result is obtained with the photon-ALP coupling $g_{a\gamma}$ from the FSRQ 4C+21.35.
However, this result shows a new photon-ALP excluded region at the low-mass ($m_a \simeq 2\times 10^{-10}\, \rm eV$) region.
Additionally, we also test the effects of the magnetic field environments model parameters $B^{\rm BLRMF}$ ($B_0^{\rm BJMF}$) and $R_{\rm BLR}$ on the final ALP limits.

The results in this paper are of great importance to the next generation VHE $\gamma$-ray measurements, like the Large High Altitude Air Shower Observatory (LHAASO) \cite{Cao:2010zz}, Tunka Advanced Instrument for Gamma-ray and Cosmic ray Astrophysics-Hundred Square km Cosmic ORigin Explorer (TAIGA-HiSCORE) \cite{Kuzmichev:2018mjq}, High Energy cosmic-Radiation Detection (HERD) \cite{Huang:2015fca}, Cherenkov Telescope Array (CTA) \cite{Acharya:2013sxa}, and Gamma-Astronomy Multifunction Modules Apparatus (GAMMA~400) \cite{Egorov:2020cmx}, which would also provide more accurate data for the FSRQ blazars to set the ALP limits.

\section*{Acknowledgments}
The author would like to thank Ciaran A.J. O'Hare for his public webpage \url{https://github.com/cajohare/AxionLimits} \cite{ciaran_o_hare_2020_3932430} from which a number of photon-ALP limits are given.
This work is supported by the National Natural Science Foundation (NNSF) of China (Grants No.~11775025 and No.~12175027).

\begin{appendix}
\section{Statistic method}
\label{app_1}

In this section, we introduce the statistic method to set the 95\% $\rm C.L.$ ALP limit in the $m_a-g_{a\gamma}$ plane. 
The test statistic (TS) can be defined as \cite{TheFermi-LAT:2016zue, Li:2020pcn} 
\begin{eqnarray}
{\rm TS} ={\widehat{\chi}_{\rm null}}^2 - {\widehat{\chi}_{\rm ALP}}^2, 
\end{eqnarray}
where ${\widehat{\chi}_{\rm null}}^2$ and ${\widehat{\chi}_{\rm ALP}}^2$ are the best-fit ${\chi}^2$ under the null and ALP hypotheses for each measurement set in the Monte Carlo simulations, respectively. 
For all 400 data sets, the TS distribution can be described by the non-central $\chi^2$ distribution with the effective degree of freedom ($\rm d.o.f.$) and the non-centrality $\lambda$. 
We assume that the above TS distribution is approximated with the ALP hypothesis and use it to obtain the value of $\Delta\chi_{95\%}^2$ at 95\% $\rm C.L.$
Then we can derive the threshold value $\chi_{95\%}^2$ to set the 95\% $\rm C.L.$ ALP limit on the ALP parameter space with 
\begin{eqnarray}
\chi_{95\%}^2 = \chi_{\rm min}^2 + \Delta\chi_{95\%}^2,
\end{eqnarray}
where ${\chi}_{\rm min}^2$ corresponds to the minimum best-fit parameter set under the ALP hypothesis.

Here we briefly introduce the non-central $\chi^2$ distribution  \cite{Cowan:2010js}.
Generally, we have the $\chi^2$ density function with the degree $\nu$ 
\begin{eqnarray}
f(x,\nu)=\frac{1}{2^{\frac{\nu}{2}}\varGamma(\frac{\nu}{2})}x^{\frac{\nu}{2}-1}e^{-\frac{x}{2}},
\end{eqnarray}
with the celebrated Gamma function 
\begin{eqnarray}
\varGamma(a)=\int_0^{\infty}t^{a-1}e^{-t}{\rm d}t.
\end{eqnarray}
Then the non-central $\chi^2$ distribution with the non-centrality $\lambda$ can be described by the Poisson mixture of the $\chi^2$ density function
\begin{eqnarray}
f(x,\nu,\lambda)=\sum_{j=0}^\infty \frac{e^{-\frac{\lambda}{2}}({\frac{\lambda}{2}})^j}{j!}f(x,\nu+2j).
\end{eqnarray}
The cumulative distribution function (CDF) can be described by
\begin{eqnarray}
P(x,\nu,\lambda)=\int_0^x f(t,\nu,\lambda){\rm d}t.
\end{eqnarray}

\end{appendix}

\bibliography{references}
\end{document}